\documentclass[twocolumn,showpacs,preprintnumbers,amsmath,amssymb,pre]{revtex4}

\usepackage{graphicx}
\usepackage{dcolumn}
\usepackage{bm}
\usepackage{subeqnar}
\usepackage{tabularx}
\usepackage{epstopdf,overpic,color}

\graphicspath{{}}

\newcommand{\bx}{\mathbf{x}}
\newcommand{\bF}{\mathbf{F}}
\newcommand{\bA}{\mathbf{A}}
\newcommand{\bX}{\mathbf{X}}
\newcommand{\by}{\mathbf{y}}
\newcommand{\bg}{\mathbf{g}}
\newcommand{\bphi}{\boldsymbol{\phi}}
\newcommand{\bPhi}{\boldsymbol{\Phi}}

\DeclareMathAlphabet{\mathpzc}{OT1}{pzc}{m}{it}
\def\bAc{{{{\boldsymbol{{\mathpzc{{A}}}}}}}}

\begin{document}

\title{Dynamic Mode Decomposition and Sparse Measurements for Characterization and Monitoring of Power System Disturbances}
\author{J. Jorge Ramos$^1$
and J. Nathan Kutz$^2$\footnote{Electronic address: \texttt{kutz@uw.edu}}}
%
\affiliation{$^1$ Graduate Program in Electrical Engineering, Cinvestav Unidad Guadalajara, Guadalajaram Jal, Mexico}
\affiliation{$^2$ Department of Applied Mathematics, University of Washington, Seattle, WA. 98195}
\date{\today}
\begin{abstract}
We introduce the dynamics mode decomposition for monitoring wide-area power grid networks from sparse measurement data.
The mathematical framework fuses data from multiple sensors based on multivariate statistics, providing accurate full state estimation from limited measurements and generating data-driven forecasts for the state of the system.  Our proposed data-driven strategy, which is based on energy metrics, can be used for the analysis of major disturbances in the network. The approach is tested and validated using time domain simulations in the IEEE 118 bus system under various disturbance scenarios and under  different sparse observations of the system.  In addition to state reconstruction, the minimal number of sensors required for monitoring disturbances can be evaluated.  Visualization techniques are developed in order to aid in the analysis and characterization of the system after disturbance. 
\end{abstract}

\maketitle

\section{Introduction}
Advanced wide-area monitoring of power grid systems that can continuously assess the power system health and performance are highly desirable~\cite{messina2015wide}. Central to such monitoring schemes are intelligent sensing methods, signal processing and communication technologies to make optimal use of measured wide-area data~\cite{kezunovic2013role}.  Intelligent monitoring devices collect measurements form Phasor Measurement Units (PMUs) and other devices, convert the incoming measurements into useful information, display this information to operators, store and analyze data to track system health, and trigger alarms whenever monitoring systems detect an abnormality~\cite{barocio2013detection}, such as the emergence of the {\em coherent swing instability} (CSI)~\cite{susuki2009global,susuki2011coherent,susuki2011nonlinear,susuki2012nonlinear,susuki2014nonlinear}.   However, while sensors have increased the quality and availability of data, the utility of algorithms that are central to timely detection and display of adverse conditions have continued to advance at a slower pace.  Towards this end, we develop a monitoring algorithm enabled by the recently developed {\em dynamic mode decomposition} (DMD) that allows for state-space reconstruction and forecasting with limited measurements of the power grid system, thus enabling real-time monitoring capabilities.  In contrast to recent DMD innovations for characterizing CSI in powergrids ~\cite{susuki2009global,susuki2011coherent,susuki2011nonlinear,susuki2012nonlinear,susuki2014nonlinear}, our focus is on optimal and sparse monitoring of the powergrid system.  Specifically, we consider which nodes, and how few, are required to be monitored in order produce accurate assessments of power grid disturbances.

Among recently developed algorithms, event location strategies have been gaining increasing attention. These strategies are geared towards providing a system-wide awareness of events such as faults and other disturbances, taking advantage of the increasing coverage of {\em wide area measurement systems} (WAMS) technology, and enabling the implementation of wide area emergency and restorative control applications~\cite{li2010online,mei2008clustering,bhui2016application}. Some of these approaches use practical assumptions about the system, such as the homogeneity of the fault propagation speed, in order to triangulate the disturbance location using frequency measurements~\cite{li2010online}.  However, these assumptions are not realistic for all operational conditions.  In addition, clustering algorithms have also been proposed in order to identify the location of the disturbance by comparing simulation cases with real measurements~\cite{mei2008clustering}.  Both approaches require accurate models and extensive simulation for offline training. More recently, recurrence quantification analysis was used to analyze voltage signals and locate disturbances~\cite{bhui2016application}, where the impact of noise and low observability are mentioned as interesting aspects to be taken in to account in the application of disturbance detection algorithms.  

Although these strategies have achieved positive results, additional technical challenges arise as the modern WAMS-generated data becomes high dimensional and more distributed through larger areas of the system.  Thus a key feature for future situational awareness schemes will be the ability to combine and analyze high dimensional and distributed information provided by several regional control systems\cite{messina2015wide}, considering the compromise between the reduction of data dimensionality and the communication capacity~\cite{kezunovic2013role}. 
These distributed wide-wide area monitoring approaches offer advantages in terms of financial resources, communication, latency, reliability and security compared with centralized counterparts~\cite{shahraeini2011comparison}.  Additionally, they allow the use of distributed processing strategies to alleviate the burden and complexity of processing the high-dimensional data in a centralized fashion~\cite{shahraeini2011comparison,wang2015decentralized}. Wide-area data fusion architectures that can be applied for hierarchical and distributed systems have started to emerge in order to surmount these challenges, including applications ranging from mode-meter algorithms~\cite{ning2013oscillation,khalid2015improved,nabavi2015distributed} and coordinated adaptive control~\cite{eriksson2011wide,liu2015armax}.  These are only some examples of methodologies that have begun to migrate toward configurations more suitable for WAMS.

\begin{figure*}[t]
 \center
  \includegraphics[width=0.8\textwidth]{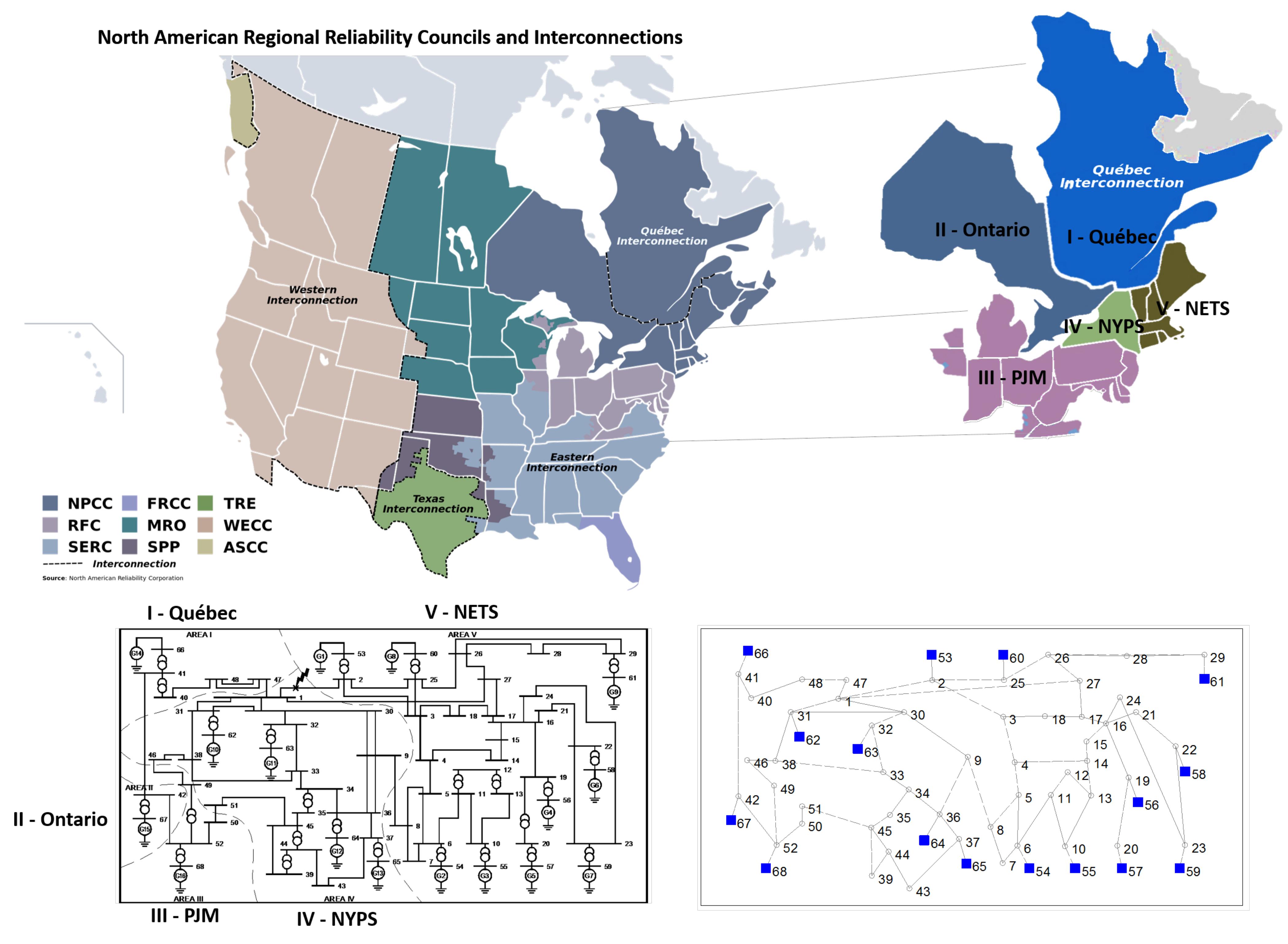}
  \caption{The IEEE 16-machine 68-bus Test System. The upper-right figure represents the 
The four NERC Interconnections, and the eight NERC Regional Reliability Organizations.  The five areas corresponding to the IIEE test system are represented geographically in the upper-left figure. The line diagram and graph of IEEE test system are shown in the bottom-left and bottom-right panels respectively.  In what follows, we will present the results of the power grid dynamics on the graph structure of the bottom right panel. }
  \vspace{-.15in}
  \label{fig2}
\end{figure*}

\begin{figure*}[t]
 \center
  \includegraphics[width=0.9\textwidth]{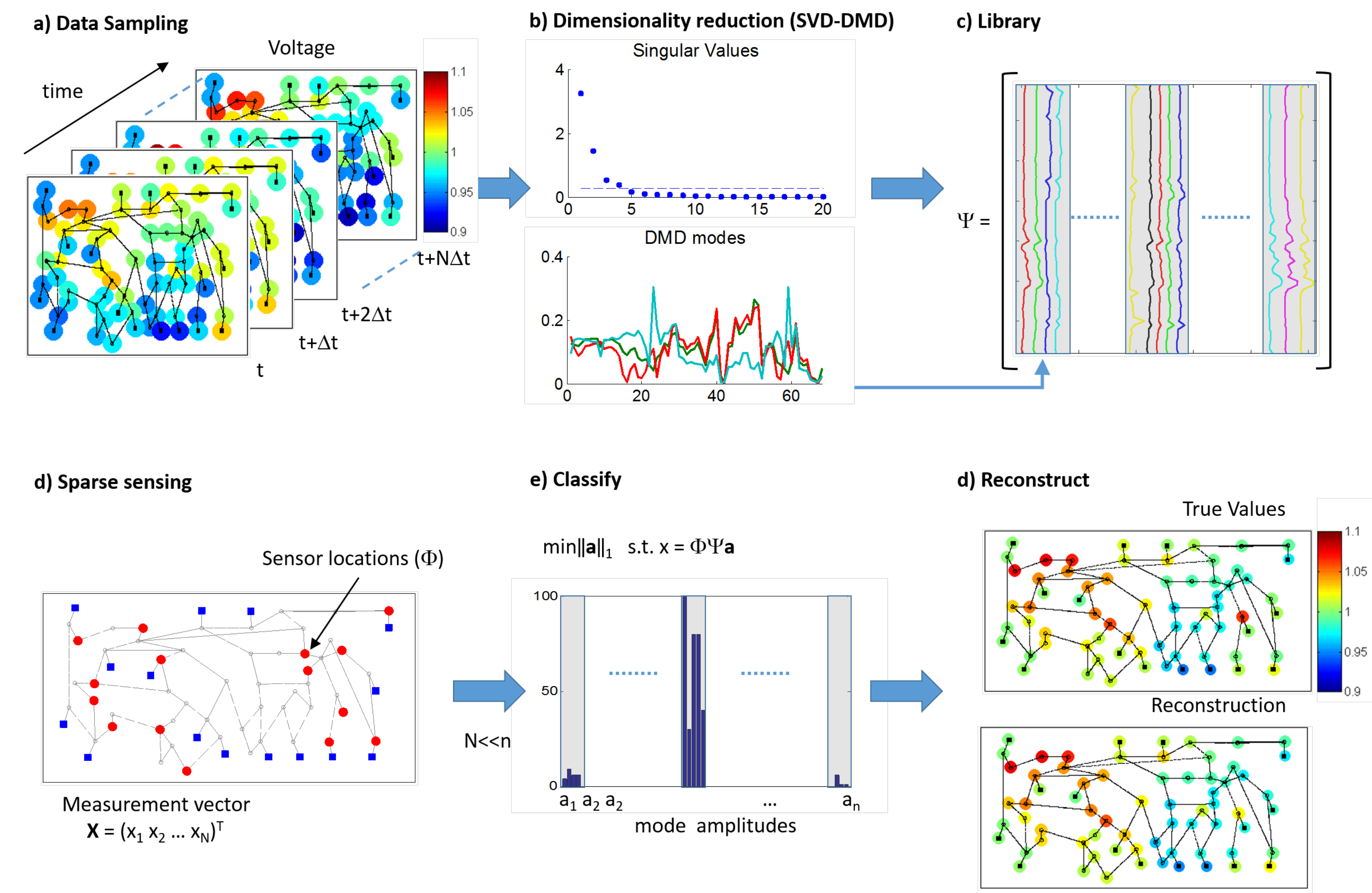}
  \caption{Algorithmic procedure for training and execution.  During the training stage (a)-(c), libraries of low-rank, spatio-temporal features are constructed from the dominant modes of the DMD algorithm.  These are stored in a library matrix ${\bf \Psi}$. The execution stages (d)-(f) show that a sparse number of sampling points (d) is required to classify the dynamic parameter regime of the system via compressive sensing (e).  The full state can then be reconstructed and a future state prediction generated (f).}
  \vspace{-.15in}
  \label{fig1}
\end{figure*}

In this paper, we consider the implementation of a disturbance detection algorithm on a distributed architecture.  Specifically, the IEEE 16-machine 68-bus Test System of the Northeastern United States (See Fig.~\ref{fig2}) is explored.  We advocate the use of the  DMD algorithm for computing interpretable, low-rank decompositions of the spatial-temporal measurements on a power grid network.  DMD has already been shown to be successful in characterizing the emergence of the deleterious CSI~\cite{susuki2009global,susuki2011coherent,susuki2011nonlinear,susuki2012nonlinear,susuki2014nonlinear}.  Our objective is quite different, but builds upon the initial success of DMD applied to powegrids.   Specifically,  by constructing libraries of dynamic activity, a small number of sensors can be used to exploit compressive sensing (sparse sampling) algorithms for classification and reconstruction of the power grid from a small number of measurements.  Figure~\ref{fig1} demonstrates the algorithmic structure proposed, which is composed of a training and execution stage.  Thus a semi-distributed tool for monitoring measured data from WAMS is presented.   A strategy for fusing data from multiple sensors into a consensus based on multivariate statistical tools (multiblock model) is first introduced and a model of the collected data is developed. Derived from this model, a data-driven strategy based on energy metrics is proposed for the location and analysis of major disturbances. Unlike the reviewed methods, the aim of the proposed strategy is to exploit the distributed processing capacity of modern WAMS architectures to visualize the system dynamic behavior following disturbances. 

The paper is structured as follows. The methodology is presented two parts: Section 2 describes the semi-distributed data fusion strategy along with its interpretation in terms of the WAMS structure, and Section 3 uses these concepts in order to derive the disturbance location strategy based on the proposed energy-based metrics. The study case and results are presented in Section 4, with the performance of the proposed method discussed and interesting remarks about the relation of disturbance location and optimal sensor placement given.  To conclude, Section 5 introduces the final discussion and future work, outlining how the DMD algorithm presents an architecture for an efficient WAMS.

%

\section{Background and DMD Algorithm}

Matrix decompositions are critically enabling algorithms for diagnostic analysis and scientific computing applications across every field of the engineering, social, biological, and physical sciences.  A primary purpose of such decompositions is the discovery of low-rank subspaces which allow one to represent the dynamics of the system of interest in an optimal way.  Of particular importance is the {\em  singular value decomposition} (SVD), which provides a principled method for dimensionality reduction and computation of interpretable subspaces within which the data resides.  So widespread is the usage of the SVD algorithm, and minor modifications thereof, that it has generated a myriad of names across various communities, including Principal Component Analysis (PCA)~\cite{Pearson:1901}, the Karhunen-Lo\`eve (KL) decomposition, Hotelling transform~\cite{hotellingJEdPsy33_1,hotellingJEdPsy33_2}, Empirical Orthogonal Functions (EOFs)~\cite{eof1} and Proper Orthogonal Decomposition 
(POD)~\cite{Berkooz:1993,HLBR_turb}.  In each of these cases, the low-rank features extracted from the matrix factorization help provide interpretable, spatially correlated structures that can help inform understanding and potential control protocols, which are particularly important in network level power grid dynamics.

Dimensionality reduction also enables the use of a small number of sensors for characterizing the dynamics (See~\cite{manohar2017data} and referenes therein).  Specifically, having computed the low-rank subspaces on which the dynamics occurs, only a sparse number of measurements are required for classification of the dynamical regime and full state reconstruction.  This has been exploited extensively in the reduced order modeling (ROM) community under the moniker of gappy POD methods~\cite{sirovich1995,willcox2005,Yildirim:2009,sorensen2010,sargsyan2015}.  For power grid applications, this allows us to monitor the health of the power system from measurements of only a small number of locations.   More precisely, the SVD provides global powergrid modes that can be exploited for improved reconstruction and prediction.

\subsection{Dynamic Mode Decomposition}

DMD is a matrix factorization method based upon the SVD algorithm.  However, in addition to performing a low-rank  approximation, it further performs and eigendecomposition on the computed subspaces in order to extract critical temporal features.  Thus the DMD method provides a spatio-temporal decomposition of data into a set of dynamic modes that are derived from snapshots or measurements of a given system in time.   The mathematics underlying the extraction of dynamic information from time-resolved snapshots  is closely related to the idea of the Arnoldi algorithm~\cite{Schmid2010jfm}, one of the workhorses of fast computational solvers.   The data collection process involves two parameters:
\begin{subeqnarray}
 &&  n = \mbox{number of spatial points saved per time snapshot} \nonumber \\
 &&  m= \mbox{number of snapshots taken} \nonumber
\end{subeqnarray}
The DMD algorithm was originally designed to collect data at regularly spaced intervals of time.  However, new innovations allow for both sparse spatial~\cite{Brunton2015jcd} and temporal~\cite{Tu:2014b} collection of data as well as irregularly spaced collection times~\cite{askham2018variable}.   Indeed, Tu \emph{et al.}~\cite{Tu2014jcd} provides a highly intuitive definition of the DMD method and algorithm.\\

\noindent {\bf Definition:  Dynamic Mode Decomposition} (Tu \emph{et al.} 2014~\cite{Tu2014jcd}): {\em Suppose we have
a dynamical system and two sets of data 
\begin{subeqnarray}
&& {\bf X} = \begin{bmatrix}
\vline & \vline & & \vline \\
\bx_1 & \bx_2 & \cdots & \bx_{m-1}\\
\vline & \vline & & \vline
\end{bmatrix} \\
&&\nonumber\\
&& {\bf X}' = \begin{bmatrix}
\vline & \vline & & \vline \\
\bx'_1 & \bx'_2 & \cdots & \bx'_{m-1}\\
\vline & \vline & & \vline
\end{bmatrix}
\end{subeqnarray}
so that $\bx'_k=\bF(\bx_k)$ where $\bF$ is the flow map corresponding to the evolution of our dynamical system for time $\Delta t$.   DMD computes the leading eigendecomposition of the best-fit linear operator $\bA$ 
relating the data $\bX'\approx\bA\bX$\,:
\begin{equation}
 {\bf A} = {\bf X}' {\bf X}^\dag.
 \label{eq:DMD}
\end{equation}
The DMD modes, also called dynamic modes, are the eigenvectors of $\bA$, and each DMD mode corresponds to a particular eigenvalue of $\bA$.}\\

In the DMD architecture, we typically consider data collected from a dynamical system
\begin{equation}
 \frac{d{\bf x}}{dt} = {\bf f}({\bf x},t;{\boldsymbol{\mu}}) \, ,
 \label{eq:U}
\end{equation}
where ${\bf x}(t)\in\mathbb{R}^n$ is a vector representing the state of our dynamical system at time $t$, $\boldsymbol{\mu}$ contains parameters of the system, and ${\bf f}(\cdot)$ represents the dynamics.  For instance, the vector ${\bf x}$ denotes the
power grid state after numerical discretization while $\boldsymbol{\mu}$ is a parametrization of the system.
The state $\bx$ is typically quite large, having dimension $n\gg 1$.  

Measurements of the system 
\begin{eqnarray}
\by_k = \bg(\bx_k),
\end{eqnarray}
are collected at times $t_k$ from $k=1,2,\cdots,m$ for a total of $m$ measurement times.  
The measurements are typically the state of the power grid, so that $\by_k=\bx_k$, however, the DMD architecture
allows for a more nuianced viewpoint of observables.  This is beyond the scope of the current work, but such ideas
are related to Koopman theory~\cite{mezic2013analysis,DMDbook}.

The DMD framework takes an equation-free perspective where the original, nonlinear dynamics (e.g. network level power grid dynamics) may be unknown.  Thus data measurements of the system alone are used to approximate the dynamics and predict the future state.  Measurements can also be made on functions of the state space, resulting in the so-called Koopman operator~\cite{mezic2013analysis,DMDbook}, which has been used previously to study power grid dynamics~\cite{susuki2009global,susuki2011coherent,susuki2011nonlinear,susuki2012nonlinear,susuki2014nonlinear}.    The DMD procedure  constructs the proxy, approximate locally linear dynamical system
\begin{equation}
 \frac{d {\bf x}}{dt} = {\large\bAc}{\bf x}\label{eq:SSLinCont}
\end{equation}
with initial condition ${\bf x}(0)$ whose well-known solution is
\begin{equation}
  {\bf x}(t)=  \sum_{k=1}^n  \bphi_k \exp(\omega_k t) b_k=\bPhi \exp(\boldsymbol{\Omega} t)\mathbf{b}\, 
  \label{eq:omegaj}
\end{equation}
where $\bphi_k$ and $\omega_k$ are the eigenvectors  and eigenvalues of the matrix ${\bf A}$, and the coefficients $b_k$ are the coordinates of $\bx(0)$ in the eigenvector basis.  

The DMD algorithm produces a low-rank eigen-decomposition of the matrix $\bA$ that optimally fits the measured trajectory $\bx_k$ for $k=1,2,\cdots,m$ in a least square sense so that
\begin{equation}
\min_{\bf A}
\|\bx_{k+1}-\bA\bx_k\|_2 \label{eq:snapshoterror}
\end{equation}
is minimized across all points for $k=1,2,\cdots, m-1$.  
The optimality of the approximation holds only over the sampling window where ${\bf A}$ is constructed, and the approximate solution can be used to not only make future state predictions, but also to derive dynamic modes critical for diagnostics.  Indeed, in much of the literature where DMD is applied, it is primarily used as a diagnostic tool.  This is much like POD analysis where the POD modes are also primarily used for diagnostic purposes.  Thus the DMD algorithm can be thought of as a modification of the SVD architecture which attempts to account for dynamic activity of the data.  The eigendecomposition of the low rank space found from SVD enforces a Fourier mode time expansion which allows one to then make spatio-temporal correlations with the sampled data.

\subsection{Compressive Sensing and Sparse Sensors}

Although the gappy POD method~\cite{sirovich1995,willcox2005,Yildirim:2009,sorensen2010,sargsyan2015} can be used for reconstruction of the full state from a small number of measurements, it does not serve well to classify the
dynamical regime given a potential number of low-rank subspaces.  Instead, we will use the compressive sensing (CS) architecture for classification of the appropriate dynamical regime~\cite{bright2013,Brunton2014siads,Proctor2014epj,kramer}.  
Once determined, a gappy reconstruction can then be performed using the modes from the selected dynamical regime.
In CS a signal that is sparse in some basis may be recovered using proportionally few measurements by solving for the $\ell_1$-minimizing solution to an underdetermined system.   

Consider a high-dimensional measurement vector of the power grid system ${\bf x}\in\mathbb{R}^n$, which is sparse in some space, spanned by the columns of a matrix ${\bf \Psi}$:
\begin{equation}
{\bf x} = {\bf \Psi a}.
\label{eq:sparse1}
\end{equation}
Here, sparsity means that $\bf x$ may be represented in the transform basis $\bf\Psi$ by a vector of coefficients $\bf a$ that contains mostly zeros.  
More specifically, $K$-\textit{sparsity} means that there are $K$ nonzero elements.  
In this sense, sparsity implies that the signal is \emph{compressible}.      

Consider a sparse measurement of the power grid system ${\bf y}\in\mathbb{R}^p$, with $p\ll n$:
\begin{equation}
{\bf y} = {\bf \Phi x},
\label{eq:sparse2}
\end{equation}
where $\Phi$ is a measurement matrix that maps the full state
measurement ${\bf x}$ to the sparse measurement vector ${\bf y}$.  Details
of this measurement matrix will be given shortly.
Plugging \eqref{eq:sparse1} into \eqref{eq:sparse2} yields an underdetermined system:
\begin{equation}
{\bf y} = {\bf \Phi\Psi a}.
\label{eq:sparse3}
\end{equation}

We may then solve for the sparsest solution ${\bf a}$ to the underdetermined system of equations in \eqref{eq:sparse3}.  Sparsity is measured by the $\ell_0$ norm, and solving for the solution ${\bf a}$ that has the smallest $|{\bf a}|_0$ norm is a combinatorially hard problem.  However, this problem may be relaxed to a convex problem, whereby the $|{\bf a}|_1$ norm is minimized, which may be solved in polynomial time~\cite{Candes:2006c,Donoho:2006}.  The specific minimization problem is:
\begin{equation*}
\arg \min {|{\bf \hat a}|_1} \text{ such that } {\bf \Phi\Psi\hat a}={\bf y}.
\end{equation*}
There are other algorithms that result in sparse solution vectors, such as orthogonal matching pursuit~\cite{tropp2007signal}.

This procedure, known as {\em compressive sensing}, is a recent development that has had widespread success across a range of problems.  There are technical issues that must be addressed.  For example, the number of measurements $p$ in ${\bf y}$ should be on the order of $K\log(n/K)$, where $K$ is the degree of sparsity of $\bf a$ in $\bf \Psi$~\cite{Candes:2006, Candes:2006a,Baraniuk:2007}.  In addition, the measurement matrix $\bf \Phi$ must be {\em incoherent} with respect to the sparse basis $\bf \Psi$, meaning that the columns of $\bf \Phi$ and the columns of $\bf \Psi$ are uncorrelated.  
Interestingly, significant work has gone into demonstrating that Bernouli and Gaussian random measurement matrices are almost certainly incoherent with respect to a given basis~\cite{Candes:2006b}. 

Typically a generic basis such as Fourier or wavelets is used in conjunction with sparse measurements consisting of random projections of the state.  However, in many engineering applications, it is unclear how random projections may be obtained without first starting with a dense measurement of the state.  In this work, we constrain the measurements to be point measurements of the state, so that $\bf \Phi$ consists of rows of a permutation matrix.  
Our primary motivation for such point measurements arises from physical considerations in such applications as ocean or atmospheric monitoring where point measurements are physically relevant.  Moreover, sparse sensing is highly desirable as each measurement device is often prohibitively expensive, thus motivating much of our efforts in using sparse measurements to characterize the complex dynamics.

\subsection{Combining Methodologies}

Figure~\ref{fig1} shows how the dimensionality reduction framework is combined with the compressive
sensing architecture.  Specifically, simulations of the the IEEE 16-machine 68-bus Test System are used to
construct low-rank embeddings of the spatio-temporal activity in the Northeast United States.  Dominant DMD modes
in various dynamic regimes are collected into a library of subspace embeddings via the matrix ${\bf \Psi}$.  Since
a given dynamic regime only requires a sparse number of modes from ${\bf \Psi}$, compressive sensing can
be used to select the correct library elements for reconstruction and future state prediction.  Thus the dimensionality
reduction and sparse sampling partner naturally for characterizing the power grid dynamics.

\section{Power Systems Model}

\begin{table}
  \begin{tabular}  {ccc}
      Area     &    Bus number    &   Total buses  \\ \hline
        1 &  1-33, 113, 114, 115, 117  &  $m_1=37$ \\
        2 & 34-76, 116, 118  & $m_2 = 45$ \\
        3 & 77-112  & $m_3 = 36$ \\ \hline
           &              &    Total: $m_T=118$
  \end{tabular}
\caption{Areas for the IEEE 118 Bus System}
\end{table}

To demonstrate the application of the proposed methodology on complex dynamical systems, we study the voltage profile of an IEEE test system model of a power system under disturbances. The IEEE test system is comprised of a 68-Bus, 16-machine, 5-Area System widely know in the power system literature. The IEEE 16-machine, 68-Bus test system is a reduced order equivalent of the interconnected New England test system (NETS) and New York power system (NYPS), with five geographical regions out of which NETS and NYPS are represented by a group of generators.  The power imported from each of the three other neighboring areas (Quebec, Ontario and PJM) are approximated by equivalent generator models as shown in Figure 2. The NETS is represented by nine generators G1 to G9 and comprises the power system of the states of Maine, New Hampshire, Vermont, Massachuset, Connecticut and Rhode Island. The NYPS corresponds to power system of New York State and it is modeled by four generators G10 to G13. Generators G14 to G16 corresponds to the I-Quebec, II- Ontario and III-PJM regions.

\begin{figure}[t]
 \hspace*{-.2in}
  \includegraphics[width=0.55\textwidth]{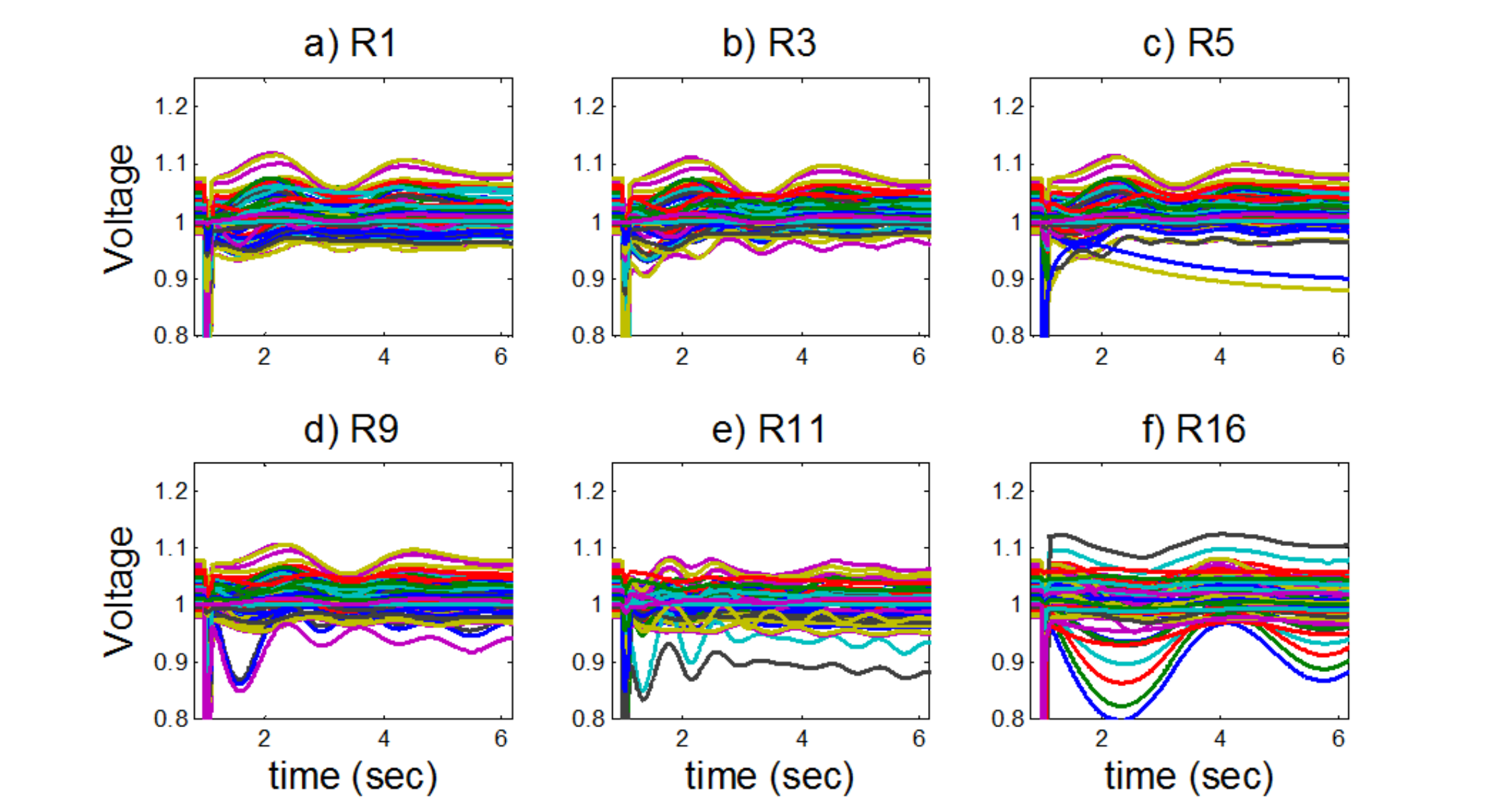}
  \caption{Voltage profile of 68 buses at different regimes. After t=1 s there is a stable state, from t = 1 to t=1.01 the fault is applied (drop of voltage), and after t=1.01 the fault is cleared and the transient measurement is presented. }
  \vspace{-.15in}
  \label{fig3}
\end{figure}

\begin{figure*}[t]
  \includegraphics[width=0.85\textwidth]{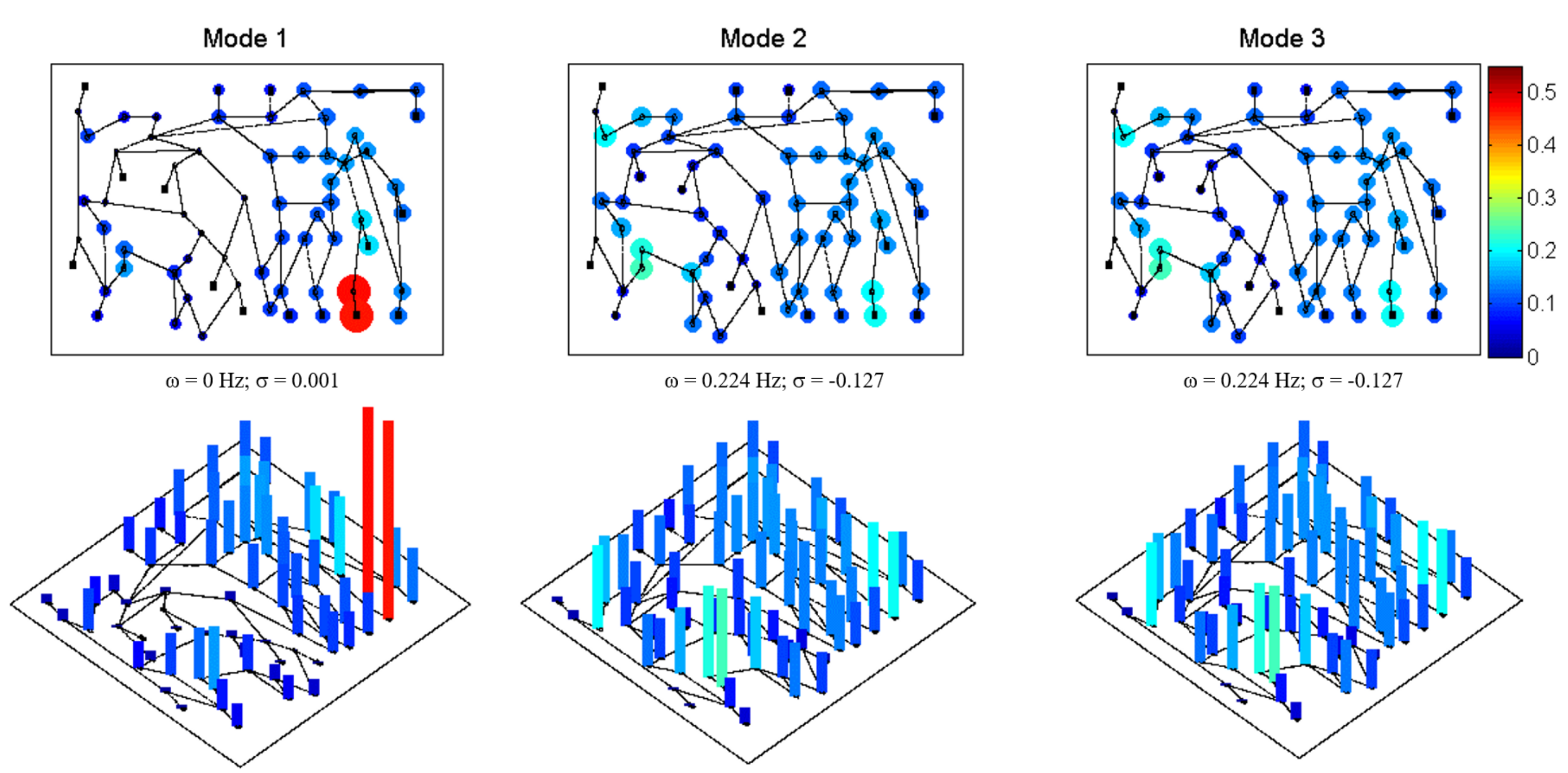}
  \caption{DMD modes representation of Regime 5 with threshold of 80\% of energy. The Mode 1 represents an exponential component with low damping, and higher values in buses close to the fault. Modes 2 and 3 represents an oscillatory mode with ? = 0.224 Hz and   =-0.127.}
  \label{fig4}
\end{figure*}

The bus data, line data, and detailed generator data of IEEE 16-machine 68-Bus test system are given in Ref.~\cite{rogers2012power}. The test system is modeled using a subtransient reactance model for all generators and an IEEE type-II AVR for G1 to G8, meanwhile G9 is equipped with IEEE type-III AVR.
 This system is chosen because it is a well-known and is a classic power system used for load flow analysis, small-signal stability analysis and non-linear simulation of the system. In our case, we use this model for non-linear simulations generated via software of power system transient analysis. The non-linear simulations are computed from the solution of the Differential-Algebraic Equations (DAE) used to represent the system. Here we are interested in the analysis of power system behavior under one of the most severe disturbance, a three-phase fault close to the bus-generator. 
  
  
In order to test the performance of the proposed methodology, 16 disturbances were simulated in the closest bus to the generator, producing 16 distinct regimes. The disturbances, a three-phase fault at a respective bus, were introduced after one second of steady state operation and cleared after 0.1 seconds. Since the phenomenon of interest after the disruption are low frequency oscillations between 0.1 to 2 Hz, a time interval of 6 seconds is used to analyze the behavior of the system after fault clearance.  The 6 second time window is sufficiently long so that at least a half cycle of the slowest  oscillation can be captured. Typical voltage profiles of 68 buses for most representative regimes are shown in Fig. 3.
 The voltage of 68 buses were sample at a rate of 100 Hz by every regime.   We then consider the possibility of using only 10, 15, 20 and 30 buses for the identification, classification and reconstruction of the dynamic regime.   Sensor placement algorithms  are described in the next section.

\begin{figure*}[t]
  \includegraphics[width=0.85\textwidth]{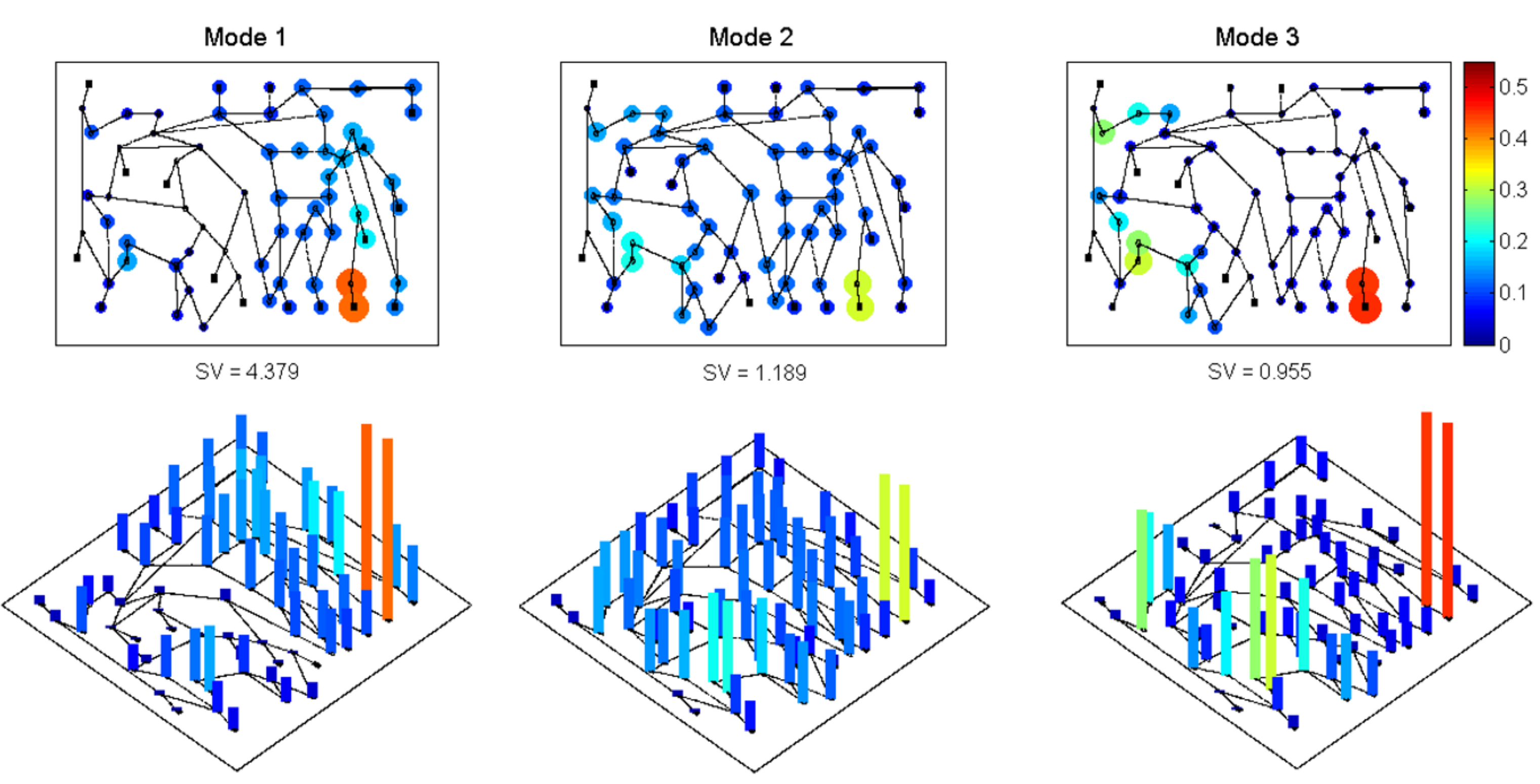}
  \caption{POD modes representation of Regime 5 with threshold of 80\% of energy. The Modes 1, 2 and 3  represents irregular oscillatory components whithin higher values in buses close to the fault.}
  \label{fig5}
\end{figure*}

\begin{figure}[t]
  \includegraphics[width=0.45\textwidth]{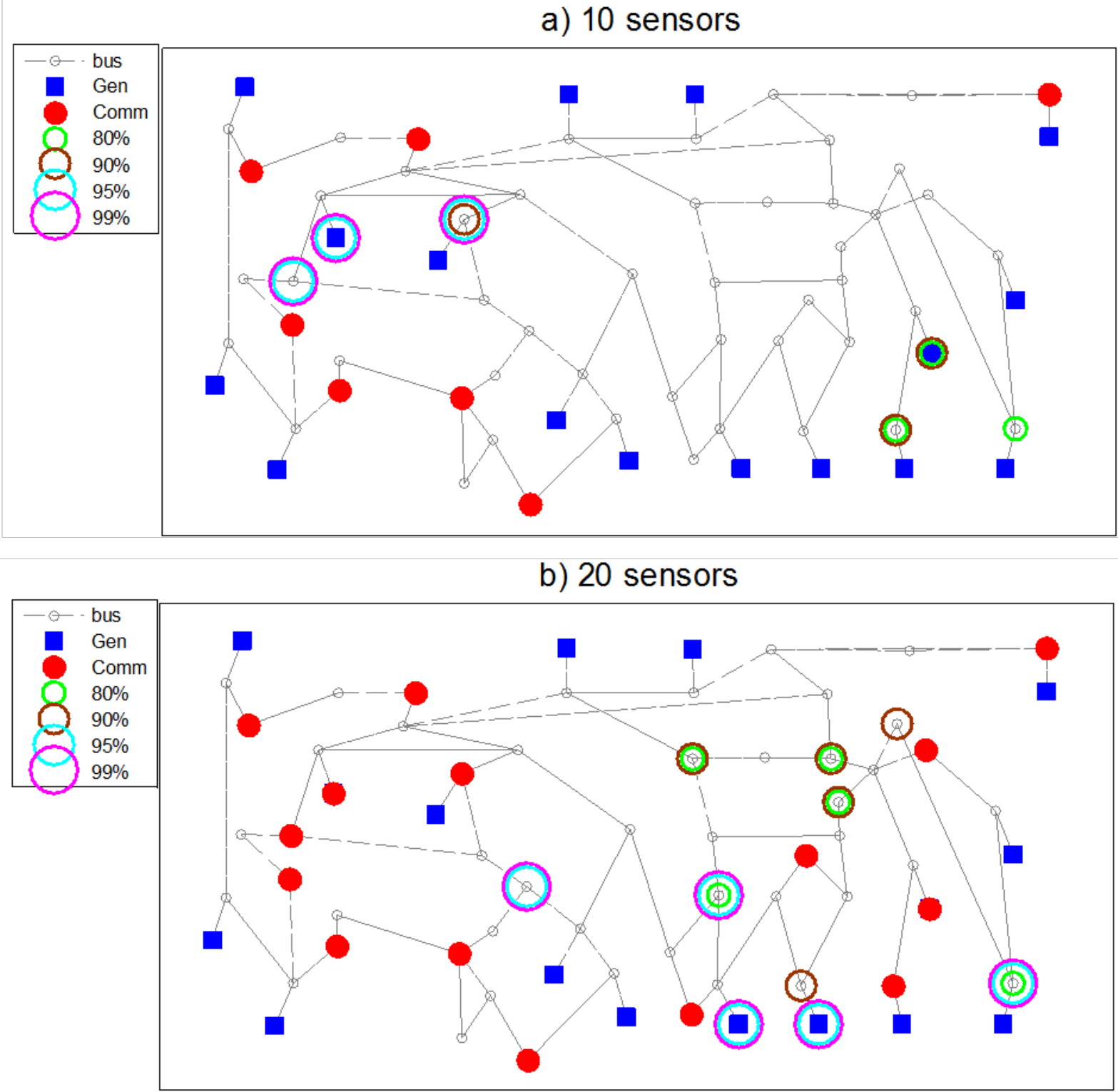}
  \caption{Graph of sparse sensor placement on IEEE 16-machine 68-bus test system at different thresholds of energy (80\%, 90\%, 95\% and 99\%) for 10 and 20 sensors. The blue square represents the generators location. The red dots indicate the common sensor between thersholds energy and color circles shows independent sensors at every threshold}
  \vspace{-.15in}
  \label{fig6}
\end{figure}

\begin{figure}[t]
  \includegraphics[width=0.45\textwidth]{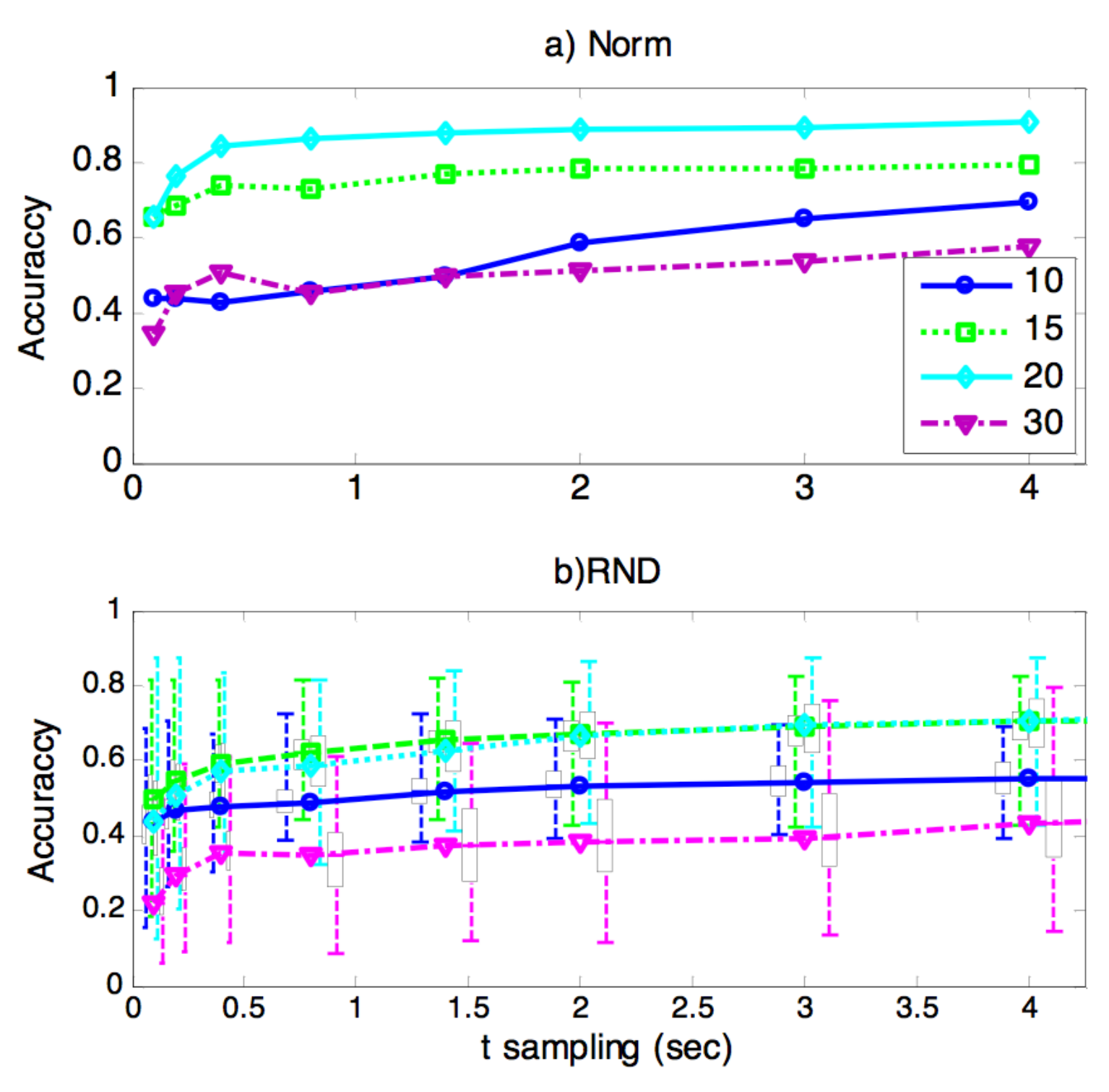}
  \caption{Accuracy of sensor placement for the library at 80\% of energy at different time intervals and number of sensors considering a) Euclidean norm and b) Random sensor placement. There are considered a window of t=[0.1, 0.2, 0.4, 0.8, 1.4, 2.0, 3.0, 4.0] s. A 100 trials of random sensor placement are considered and the results are presented by boxplot in b) RND.}
  \vspace{-.15in}
  \label{fig7}
\end{figure}

\begin{figure}[t]
  \includegraphics[width=0.45\textwidth]{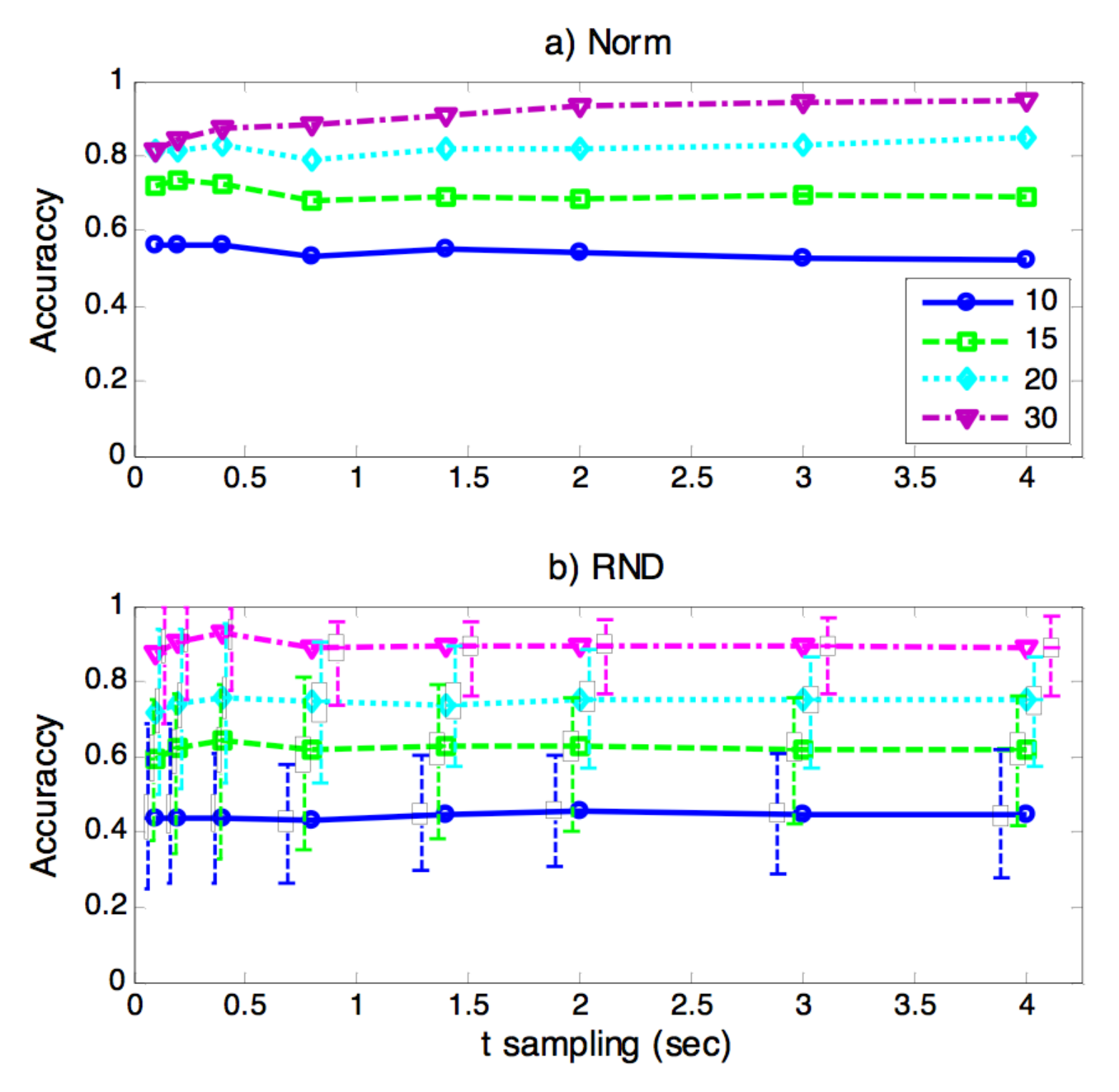}
  \caption{Accuracy of sensor placement for the library at 99\% of energy at different time intervals and number of sensors considering a) Euclidean norm and b) Random sensor placement. There are considered a window of t=[0.1, 0.2, 0.4, 0.8, 1.4, 2.0, 3.0, 4.0] s. A 100 trials of random sensor placement are considered and the results are presented by boxplot in b) RND}
  \vspace{-.15in}
  \label{fig8}
\end{figure}

\begin{figure}[b]
  \includegraphics[width=0.5\textwidth]{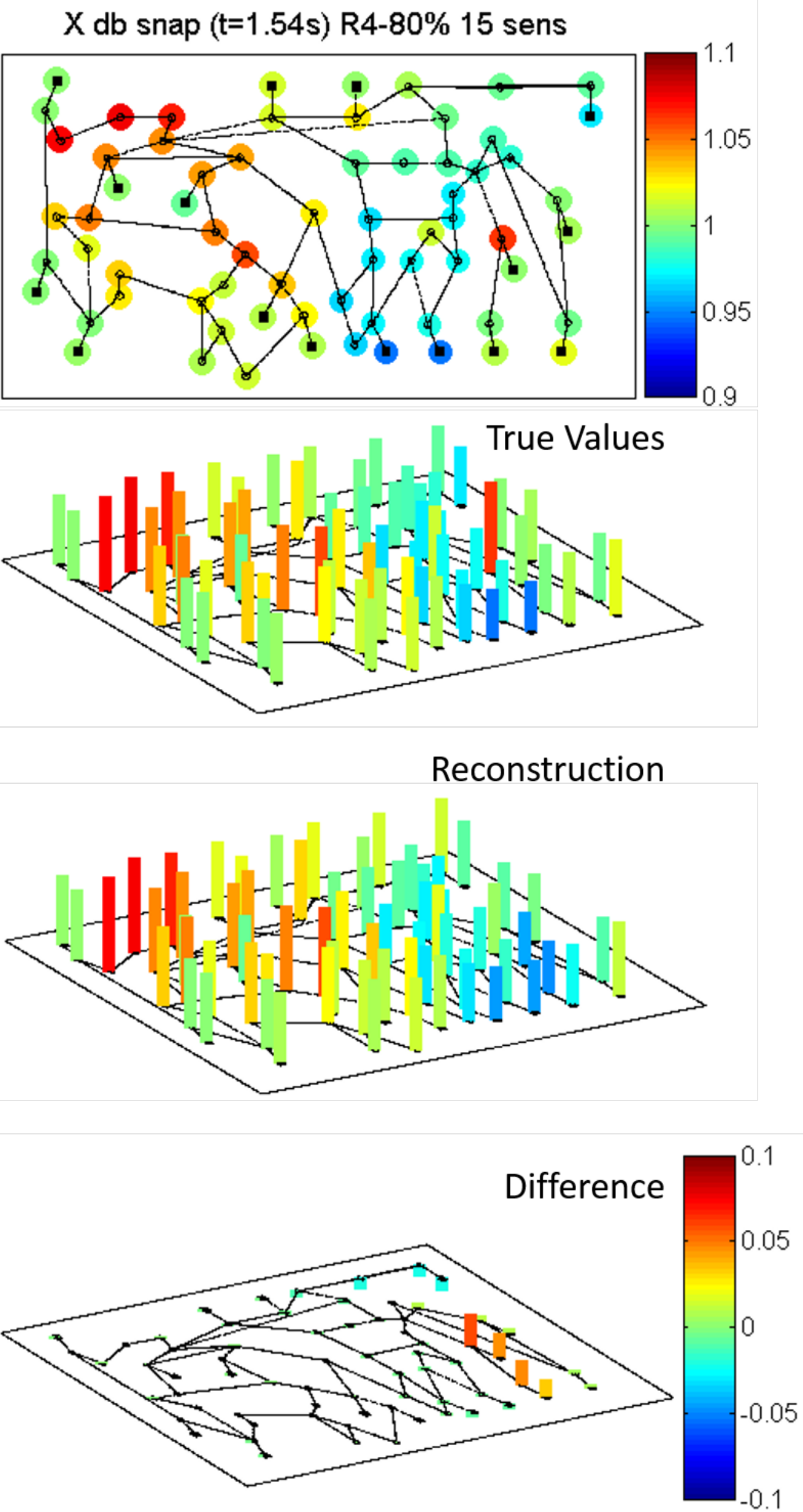}
  \caption{Voltage profile reconstruction of Regime 4 at t=1.54 s using 15 sensors and library of 80\% of energy. The upper image shows a graph of IEEE test system with voltage profiles values represented though circle?s diameter and colorbar.The middle images represents voltage profiles values as a colored bars for true values and reconstruction values. Finally the lower figure shows the difference between true values and reconstruction values.}
  \vspace{-.15in}
  \label{fig9}
\end{figure}

\begin{figure}[t]
  \includegraphics[width=0.5\textwidth]{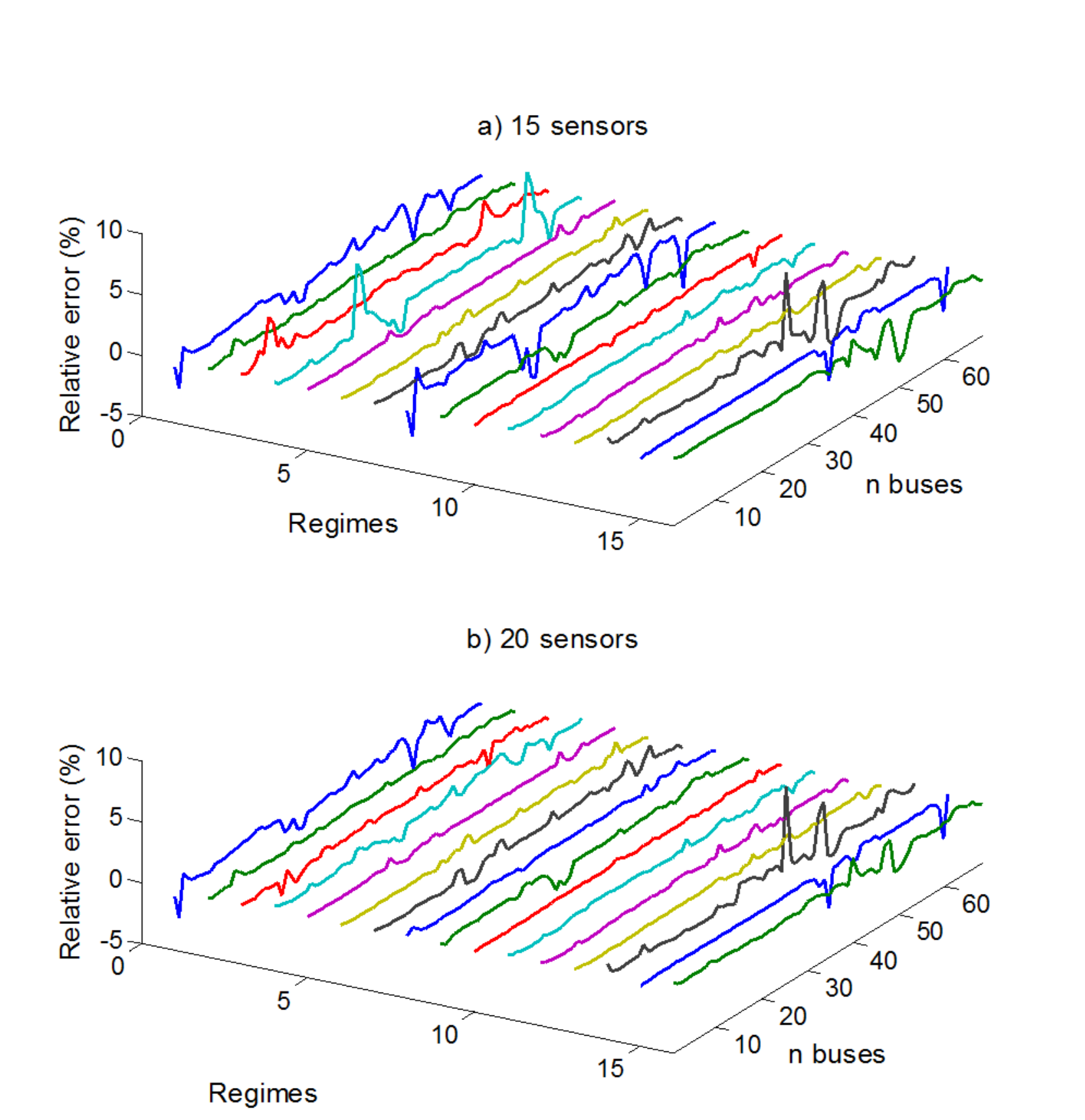}
  \caption{Relative error for full voltage profile recontruction of all regimes at t=1.54 s using 15 and 20 sensors and library of 80\% of energy.}
  \vspace{-.15in}
  \label{fig10}
\end{figure}

\section{Power Systems Networks Dynamics}
 
The power system network dynamics are evaluated in two stages. During the first stage,  a library of the dominant DMD modes all are extracted in order to determine a sensor placement. In the sparse measurement stage, the limited measurements are used to analyze, identify, classify and reconstruct the full voltages profiles of the power system.

\subsection{Library Learning} 

The 68-voltage measurements of every regime are analyzed by the DMD method to get a low dimensional representation of the data.   DMD is used to construct a library of dominant DMD modes, which are the library elements that encode the low-rank dynamics of the power system for a specific regime or fault location. 

\begin{table*}
   \begin{tabular}{c|rrrrrrrrrrrrrrrr|c}
        &    &   &    &  &  & {\bf DMD Regimes}\hspace*{-.9in}  &  &  &  &  &  &  &  &  &  &  &  \\[.1in] \hline 
       Threshold & 1  & 2  &  3  & 4 & \color{red} 5\color{black} & 6 & 7 & 8 & 9 & 10 & 11 & 12 & 13 & 14 & 15 & 16 & Library \\ 
       &   &   &    &  &  &  &  &  &  &  &  &  &  &  &  &  & elements \\ \hline 
      80\%     &  2  & 3  &  2  & 2 & \color{red} 3\color{black} & 3 & 2 & 3 & 2 & 4 & 2 &  2  &  4  & 2  &  2 & 2 & 40\\ 
        90\%  &  4  & 4  &  4  & 3 & \color{red} 4\color{black} & 4 & 3 & 5 & 3 & 6 & 4 &  4  &  5  & 3  &  3 & 3 & 62\\ 
        99\%   &  10  & 11  & 11  & 8 & \color{red} 9\color{black} & 11 & 10 & 11 & 8 & 12 & 10 &  11  &  11  & 8  &  6 & 8 & 155
   \end{tabular}\\[.3in]
   \begin{tabular}{c|rrrrrrrrrrrrrrrr|c}
        &    &   &    &  &  & {\bf POD Regimes}\hspace*{-.9in}  &  &  &  &  &  &  &  &  &  &  &  \\[.1in] \hline 
       Threshold & 1  & 2  &  3  & 4 & \color{red} 5\color{black} & 6 & 7 & 8 & 9 & 10 & 11 & 12 & 13 & 14 & 15 & 16 & Library \\ 
       &   &   &    &  &  &  &  &  &  &  &  &  &  &  &  &  & elements \\ \hline 
      80\%     &  2  & 3  &  2  & 2 & \color{red} 3\color{black} & 3 & 2 & 3 & 2 & 4 & 2 &  2  &  4  & 2  &  2 & 2 & 40\\ 
        90\%  &  4  & 4  &  4  & 3 & \color{red} 4\color{black} & 4 & 3 & 5 & 3 & 6 & 4 &  4  &  5  & 3  &  3 & 3 & 62\\ 
        99\%   &  10  & 11  & 11  & 8 & \color{red} 9\color{black} & 11 & 10 & 11 & 8 & 12 & 10 &  11  &  11  & 8  &  6 & 8 & 155
   \end{tabular}
\caption{Number of modes for every regime using DMD method and POD method for a threshold 80\%, 90\% and 99\% of energy.}
\end{table*}

The number of DMD modes choosen to be part of library are limited to those modes that captures a prescribed percentage of energy (or observed variance), based on singular values from the SVD computed during the DMD procedure. The full state voltage profiles are low-rank so that the dynamics of the power system can be represented by a sparse number of modes. These modes are called the dominant modes. Table 1 shows the number of DMD modes selected to represent every regime according to the energy threshold of 80\%, 90\% and 99\%.   A comparison of the low dimensional representation between the DMD method and the POD modes is also presented in Table I.
Noticed that even when the number of modes for both methods are the same, their representation and meaning for the power network are different as it can be seen in Fig. 4 and Fig. 5.

\subsection{Sparse Sensor Placement for Classification}

There are several ways to measure the importance of every sensor (or node) to the global behavior of the power system.   A straightforward way is to measure the Euclidean norm between elements of library to rank every sensor. These projection onto the library elements give us an average of the importance of every sensor in the different regimes.  The sensors can then be ordered from highest to lowest in the Euclidean norm, thus informing one of the number of sensors needed. The Fig. 6 shows a graph of sparse sensor placement on IEEE test system at different threshold of energy for the cases of 10 and 20 sensors.

Once the sparse sensor placement is defined, compressive sensing is used to  identify, classify and reconstruct the full voltage profile of power system.
The identification and classification of every regime using a specific number of sensor are evaluating using the principles of compressive sensing outlined previously. The main assumption of compressed sensing in this context is that given measurement vector ${\bf y}={\bf \Phi \Psi a}$ is sparsely represented in ${\bf a}$. Thus the objective is to find a basis that represents our data ${\bf y}$ in a sparse manner based on $\min \|{\bf a} \|_1$ subject to ${\bf y=\Phi \Psi a}$. Once ${\bf a}$ is defined, the highest values of ${\bf a}$ corresponds to the element of the library associated with the selected regime.
The accuracy of proposed methodology in the identification and classification of regimes using a specific number of sensor  is evaluated by averaging the identified regime of ${\bf y}$ every 5 samples until the number of samples corresponding to the window length i.e. for window length w = 0.1 s, there are 10 sample so the vector ${\bf a}$ is calculated twice. Then the average of all evaluations of a for every regime for every window length represents the accuracy of the proposed method. The results of these evaluations are shown in Fig. 7 and Fig. 8 when truncating the DMD library modes at 80\% and 99\% variance respectively.
 
In general,  better accuracy is achieved for the sparse sensor placement algorithm using the Euclidean norm versus random sensor placement. Also when the sampling window length of $t$ reaches 0.8 and 1.4 seconds, the accuracy is effectively maximized.  Thus, increasing the length of the sampling window does not improve performance.  When 15 and 20 sensors are used, the accuracy achieves values of 70 to 90\% for both libraries.
In agreement with the previous analysis, a sparse sensor placement using 15 and 20 sensors and with a sampling window length of 1.4 seconds and library constructed with  80\% variance, reconstruction of full state is achieved. This selection implies a 77 and 88\% accuracy for the case of 15 and 20 sensors respectively.
 
 \subsection{Full State Reconstrution}
 
 The full voltage profile reconstruction can be easily achieved once the classification task is accomplished. The procedure consists in projecting the data measurements onto the identified dominant modes of the library for the given regime found from classification. 
The full sate reconstruction is considered using a window length of 1.4 seconds for the identification of the regime and 15 and 20 sparse sensors placed using the library with 80\% energy variance threshold. An example of the full voltage profile reconstruction is shown in Fig. 9. This reconstruction corresponds to the worst case of all regimes and the values of the difference between the true values and reconstruction values are located in the buses close to the fault. Even when this values exist, they are meaningless.

The relative error of full voltage profile reconstruction of all regimes at a specific snapshot using 15 and 20 sensors and library of 80\% variance is displayed in Fig. 10. Noticed that higher values of relative error for 15 sensors belongs to R4, R8, R14 and R16, which suggests that four modes could not easily be correctly identified.  However, the values of relative error for most of buses of these regimes remain at low levels. These results are improved when we use 20 sensors instead of 15 sensors. In this case, there are only two regimes with errors, R14 an R16.  This results shown in Fig. 10 agrees with the accuracy results of Fig. 7. 
 
\section{Conclusions}

This paper proposes a data-driven framework for the analysis and visualization of power system disturbances. The approach is based on a semi-distributed algorithm that allows the computation of energy-based metrics from extracted consensus components.  Specifically, we have further developed the DMD algorithm for characterizing the dynamics of disturbances in power grid networks and monitoring wide-area power grid networks from sparse measurement data.
Our proposed data-driven strategy, which is based on energy metrics, can be used for the analysis of major disturbances in the network. The approach is tested and validated using time domain simulations in the IEEE 118 bus system under various disturbance scenarios and under  different sparse observations of the system.  In addition to state reconstruction, the minimal number of sensors required for monitoring disturbances can be evaluated.  Visualization techniques are developed in order to aid in the analysis and characterization of the system after disturbance. 
 
With the emergence of advanced wide-area monitoring of power grid systems, it is important to develop data-driven methods that can continuously assess the power system health and performance. Central to such monitoring schemes are intelligent sensing methods, signal processing and communication technologies to make optimal use of measured wide-area data.   We propose a new methodology for converting sparse, real-time measurements of a power grid into useful information that can be used to reconstruct the entire state space and produce short-time forecasts. The utility of such algorithms are central to timely detection and display of adverse conditions in the power grid. We have shown that the recently developed {\em dynamic mode decomposition} (DMD) is a promising data-driven method that allows for full state-space reconstruction and forecasting with limited measurements of the power grid system, thus enabling real-time monitoring capabilities.   
 
\section*{Acknowledgements}
JJR acknowledges support from National Council for Science and Technology of Mexico (CONACyT) under the grant No. 290733 and Cinvestav IPN.  JNK acknowledges support from the U.S. Air Force Office of Scientific Research (FA9550-19-1-0011).

\setlength{\bibsep}{1.75pt}
\footnotesize{\bibliographystyle{unsrt}
\bibliography{bibALL}}

\begin{thebibliography}{49}
\expandafter\ifx\csname natexlab\endcsname\relax\def\natexlab#1{#1}\fi
\expandafter\ifx\csname bibnamefont\endcsname\relax
  \def\bibnamefont#1{#1}\fi
\expandafter\ifx\csname bibfnamefont\endcsname\relax
  \def\bibfnamefont#1{#1}\fi
\expandafter\ifx\csname citenamefont\endcsname\relax
  \def\citenamefont#1{#1}\fi
\expandafter\ifx\csname url\endcsname\relax
  \def\url#1{\texttt{#1}}\fi
\expandafter\ifx\csname urlprefix\endcsname\relax\def\urlprefix{URL }\fi
\providecommand{\bibinfo}[2]{#2}
\providecommand{\eprint}[2][]{\url{#2}}

\bibitem[{\citenamefont{Messina and Messina}(2015)}]{messina2015wide}
\bibinfo{author}{\bibfnamefont{A.~R.} \bibnamefont{Messina}} \bibnamefont{and}
  \bibinfo{author}{\bibfnamefont{A.~R.} \bibnamefont{Messina}},
  \emph{\bibinfo{title}{Wide-area monitoring of interconnected power systems}}
  (\bibinfo{publisher}{The Institution of Engineering and Technology},
  \bibinfo{year}{2015}).

\bibitem[{\citenamefont{Kezunovic et~al.}(2013)\citenamefont{Kezunovic, Xie,
  and Grijalva}}]{kezunovic2013role}
\bibinfo{author}{\bibfnamefont{M.}~\bibnamefont{Kezunovic}},
  \bibinfo{author}{\bibfnamefont{L.}~\bibnamefont{Xie}}, \bibnamefont{and}
  \bibinfo{author}{\bibfnamefont{S.}~\bibnamefont{Grijalva}}, in
  \emph{\bibinfo{booktitle}{Bulk Power System Dynamics and Control-IX
  Optimization, Security and Control of the Emerging Power Grid (IREP), 2013
  IREP Symposium}} (\bibinfo{organization}{IEEE}, \bibinfo{year}{2013}), pp.
  \bibinfo{pages}{1--9}.

\bibitem[{\citenamefont{Barocio et~al.}(2013)\citenamefont{Barocio, Pal,
  Fabozzi, and Thornhill}}]{barocio2013detection}
\bibinfo{author}{\bibfnamefont{E.}~\bibnamefont{Barocio}},
  \bibinfo{author}{\bibfnamefont{B.~C.} \bibnamefont{Pal}},
  \bibinfo{author}{\bibfnamefont{D.}~\bibnamefont{Fabozzi}}, \bibnamefont{and}
  \bibinfo{author}{\bibfnamefont{N.~F.} \bibnamefont{Thornhill}}, in
  \emph{\bibinfo{booktitle}{Bulk Power System Dynamics and Control-IX
  Optimization, Security and Control of the Emerging Power Grid (IREP), 2013
  IREP Symposium}} (\bibinfo{organization}{IEEE}, \bibinfo{year}{2013}), pp.
  \bibinfo{pages}{1--10}.

\bibitem[{\citenamefont{Susuki et~al.}(2009)\citenamefont{Susuki, Mezic, and
  Hikihara}}]{susuki2009global}
\bibinfo{author}{\bibfnamefont{Y.}~\bibnamefont{Susuki}},
  \bibinfo{author}{\bibfnamefont{I.}~\bibnamefont{Mezic}}, \bibnamefont{and}
  \bibinfo{author}{\bibfnamefont{T.}~\bibnamefont{Hikihara}}, in
  \emph{\bibinfo{booktitle}{2009 American Control Conference}}
  (\bibinfo{organization}{IEEE}, \bibinfo{year}{2009}), pp.
  \bibinfo{pages}{3446--3451}.

\bibitem[{\citenamefont{Susuki et~al.}(2011)\citenamefont{Susuki, Mezi{\'c},
  and Hikihara}}]{susuki2011coherent}
\bibinfo{author}{\bibfnamefont{Y.}~\bibnamefont{Susuki}},
  \bibinfo{author}{\bibfnamefont{I.}~\bibnamefont{Mezi{\'c}}},
  \bibnamefont{and} \bibinfo{author}{\bibfnamefont{T.}~\bibnamefont{Hikihara}},
  \bibinfo{journal}{Journal of nonlinear science}
  \textbf{\bibinfo{volume}{21}}, \bibinfo{pages}{403} (\bibinfo{year}{2011}).

\bibitem[{\citenamefont{Susuki and Mezic}(2011)}]{susuki2011nonlinear}
\bibinfo{author}{\bibfnamefont{Y.}~\bibnamefont{Susuki}} \bibnamefont{and}
  \bibinfo{author}{\bibfnamefont{I.}~\bibnamefont{Mezic}},
  \bibinfo{journal}{IEEE Transactions on Power Systems}
  \textbf{\bibinfo{volume}{26}}, \bibinfo{pages}{1894} (\bibinfo{year}{2011}).

\bibitem[{\citenamefont{Susuki and Mezic}(2012)}]{susuki2012nonlinear}
\bibinfo{author}{\bibfnamefont{Y.}~\bibnamefont{Susuki}} \bibnamefont{and}
  \bibinfo{author}{\bibfnamefont{I.}~\bibnamefont{Mezic}},
  \bibinfo{journal}{IEEE Transactions on Power Systems}
  \textbf{\bibinfo{volume}{27}}, \bibinfo{pages}{1182} (\bibinfo{year}{2012}).

\bibitem[{\citenamefont{Susuki and Mezi{\'c}}(2014)}]{susuki2014nonlinear}
\bibinfo{author}{\bibfnamefont{Y.}~\bibnamefont{Susuki}} \bibnamefont{and}
  \bibinfo{author}{\bibfnamefont{I.}~\bibnamefont{Mezi{\'c}}},
  \bibinfo{journal}{IEEE Transactions on Power Systems}
  \textbf{\bibinfo{volume}{29}}, \bibinfo{pages}{899} (\bibinfo{year}{2014}).

\bibitem[{\citenamefont{Li et~al.}(2010)\citenamefont{Li, Tang, Ma, and
  Liu}}]{li2010online}
\bibinfo{author}{\bibfnamefont{W.}~\bibnamefont{Li}},
  \bibinfo{author}{\bibfnamefont{J.}~\bibnamefont{Tang}},
  \bibinfo{author}{\bibfnamefont{J.}~\bibnamefont{Ma}}, \bibnamefont{and}
  \bibinfo{author}{\bibfnamefont{Y.}~\bibnamefont{Liu}}, \bibinfo{journal}{IEEE
  Transactions on Smart Grid} \textbf{\bibinfo{volume}{1}},
  \bibinfo{pages}{253} (\bibinfo{year}{2010}).

\bibitem[{\citenamefont{Mei et~al.}(2008)\citenamefont{Mei, Rovnyak, and
  Ong}}]{mei2008clustering}
\bibinfo{author}{\bibfnamefont{K.}~\bibnamefont{Mei}},
  \bibinfo{author}{\bibfnamefont{S.~M.} \bibnamefont{Rovnyak}},
  \bibnamefont{and} \bibinfo{author}{\bibfnamefont{C.-M.} \bibnamefont{Ong}},
  \bibinfo{journal}{IEEE Transactions on Power Systems}
  \textbf{\bibinfo{volume}{23}}, \bibinfo{pages}{673} (\bibinfo{year}{2008}).

\bibitem[{\citenamefont{Bhui and Senroy}(2016)}]{bhui2016application}
\bibinfo{author}{\bibfnamefont{P.}~\bibnamefont{Bhui}} \bibnamefont{and}
  \bibinfo{author}{\bibfnamefont{N.}~\bibnamefont{Senroy}},
  \bibinfo{journal}{IEEE Transactions on Power Systems}
  \textbf{\bibinfo{volume}{31}}, \bibinfo{pages}{581} (\bibinfo{year}{2016}).

\bibitem[{\citenamefont{Shahraeini et~al.}(2011)\citenamefont{Shahraeini,
  Javidi, and Ghazizadeh}}]{shahraeini2011comparison}
\bibinfo{author}{\bibfnamefont{M.}~\bibnamefont{Shahraeini}},
  \bibinfo{author}{\bibfnamefont{M.~H.} \bibnamefont{Javidi}},
  \bibnamefont{and} \bibinfo{author}{\bibfnamefont{M.~S.}
  \bibnamefont{Ghazizadeh}}, \bibinfo{journal}{IEEE Transactions on Smart Grid}
  \textbf{\bibinfo{volume}{2}}, \bibinfo{pages}{206} (\bibinfo{year}{2011}).

\bibitem[{\citenamefont{Wang et~al.}(2015)\citenamefont{Wang, Yemula, and
  Bose}}]{wang2015decentralized}
\bibinfo{author}{\bibfnamefont{Y.}~\bibnamefont{Wang}},
  \bibinfo{author}{\bibfnamefont{P.}~\bibnamefont{Yemula}}, \bibnamefont{and}
  \bibinfo{author}{\bibfnamefont{A.}~\bibnamefont{Bose}},
  \bibinfo{journal}{IEEE Transactions on Smart Grid}
  \textbf{\bibinfo{volume}{6}}, \bibinfo{pages}{885} (\bibinfo{year}{2015}).

\bibitem[{\citenamefont{Ning et~al.}(2013)\citenamefont{Ning, Pan, and
  Venkatasubramanian}}]{ning2013oscillation}
\bibinfo{author}{\bibfnamefont{J.}~\bibnamefont{Ning}},
  \bibinfo{author}{\bibfnamefont{X.}~\bibnamefont{Pan}}, \bibnamefont{and}
  \bibinfo{author}{\bibfnamefont{V.}~\bibnamefont{Venkatasubramanian}},
  \bibinfo{journal}{IEEE Transactions on Power Systems}
  \textbf{\bibinfo{volume}{28}}, \bibinfo{pages}{1960} (\bibinfo{year}{2013}).

\bibitem[{\citenamefont{Khalid and Peng}(2015)}]{khalid2015improved}
\bibinfo{author}{\bibfnamefont{H.~M.} \bibnamefont{Khalid}} \bibnamefont{and}
  \bibinfo{author}{\bibfnamefont{J.~C.-H.} \bibnamefont{Peng}},
  \bibinfo{journal}{IEEE Transactions on Power Systems}
  \textbf{\bibinfo{volume}{30}}, \bibinfo{pages}{680} (\bibinfo{year}{2015}).

\bibitem[{\citenamefont{Nabavi et~al.}(2015)\citenamefont{Nabavi, Zhang, and
  Chakrabortty}}]{nabavi2015distributed}
\bibinfo{author}{\bibfnamefont{S.}~\bibnamefont{Nabavi}},
  \bibinfo{author}{\bibfnamefont{J.}~\bibnamefont{Zhang}}, \bibnamefont{and}
  \bibinfo{author}{\bibfnamefont{A.}~\bibnamefont{Chakrabortty}},
  \bibinfo{journal}{IEEE Transactions on Smart Grid}
  \textbf{\bibinfo{volume}{6}}, \bibinfo{pages}{2529} (\bibinfo{year}{2015}).

\bibitem[{\citenamefont{Eriksson and Soder}(2011)}]{eriksson2011wide}
\bibinfo{author}{\bibfnamefont{R.}~\bibnamefont{Eriksson}} \bibnamefont{and}
  \bibinfo{author}{\bibfnamefont{L.}~\bibnamefont{Soder}},
  \bibinfo{journal}{IEEE Transactions on Power Delivery}
  \textbf{\bibinfo{volume}{26}}, \bibinfo{pages}{988} (\bibinfo{year}{2011}).

\bibitem[{\citenamefont{Liu et~al.}(2015)\citenamefont{Liu, Zhu, Pan, Bai, Liu,
  Liu, Patel, Farantatos, and Bhatt}}]{liu2015armax}
\bibinfo{author}{\bibfnamefont{H.}~\bibnamefont{Liu}},
  \bibinfo{author}{\bibfnamefont{L.}~\bibnamefont{Zhu}},
  \bibinfo{author}{\bibfnamefont{Z.}~\bibnamefont{Pan}},
  \bibinfo{author}{\bibfnamefont{F.}~\bibnamefont{Bai}},
  \bibinfo{author}{\bibfnamefont{Y.}~\bibnamefont{Liu}},
  \bibinfo{author}{\bibfnamefont{Y.}~\bibnamefont{Liu}},
  \bibinfo{author}{\bibfnamefont{M.}~\bibnamefont{Patel}},
  \bibinfo{author}{\bibfnamefont{E.}~\bibnamefont{Farantatos}},
  \bibnamefont{and} \bibinfo{author}{\bibfnamefont{N.}~\bibnamefont{Bhatt}}
  (\bibinfo{year}{2015}).

\bibitem[{\citenamefont{Pearson}(1901)}]{Pearson:1901}
\bibinfo{author}{\bibfnamefont{K.}~\bibnamefont{Pearson}},
  \bibinfo{journal}{Philosophical Magazine} \textbf{\bibinfo{volume}{2}},
  \bibinfo{pages}{559} (\bibinfo{year}{1901}).

\bibitem[{\citenamefont{Hotelling}(1933{\natexlab{a}})}]{hotellingJEdPsy33_1}
\bibinfo{author}{\bibfnamefont{H.}~\bibnamefont{Hotelling}},
  \textbf{\bibinfo{volume}{24}}, \bibinfo{pages}{417}
  (\bibinfo{year}{1933}{\natexlab{a}}), ISSN \bibinfo{issn}{0022-0663}.

\bibitem[{\citenamefont{Hotelling}(1933{\natexlab{b}})}]{hotellingJEdPsy33_2}
\bibinfo{author}{\bibfnamefont{H.}~\bibnamefont{Hotelling}},
  \textbf{\bibinfo{volume}{24}}, \bibinfo{pages}{498}
  (\bibinfo{year}{1933}{\natexlab{b}}), ISSN \bibinfo{issn}{0022-0663}.

\bibitem[{\citenamefont{Lorenz}(1956)}]{eof1}
\bibinfo{author}{\bibfnamefont{E.~N.} \bibnamefont{Lorenz}},
  \bibinfo{journal}{Technical report, Massachusetts Institute of Technology}
  \textbf{\bibinfo{volume}{Dec.}} (\bibinfo{year}{1956}).

\bibitem[{\citenamefont{Berkooz et~al.}(1993)\citenamefont{Berkooz, Holmes, and
  Lumley}}]{Berkooz:1993}
\bibinfo{author}{\bibfnamefont{G.}~\bibnamefont{Berkooz}},
  \bibinfo{author}{\bibfnamefont{P.}~\bibnamefont{Holmes}}, \bibnamefont{and}
  \bibinfo{author}{\bibfnamefont{J.~L.} \bibnamefont{Lumley}},
  \bibinfo{journal}{Annual Review of Fluid Mechanics}
  \textbf{\bibinfo{volume}{23}}, \bibinfo{pages}{539} (\bibinfo{year}{1993}).

\bibitem[{\citenamefont{Holmes et~al.}(2012)\citenamefont{Holmes, Lumley,
  Berkooz, and Rowley}}]{HLBR_turb}
\bibinfo{author}{\bibfnamefont{P.~J.} \bibnamefont{Holmes}},
  \bibinfo{author}{\bibfnamefont{J.~L.} \bibnamefont{Lumley}},
  \bibinfo{author}{\bibfnamefont{G.}~\bibnamefont{Berkooz}}, \bibnamefont{and}
  \bibinfo{author}{\bibfnamefont{C.~W.} \bibnamefont{Rowley}},
  \emph{\bibinfo{title}{Turbulence, coherent structures, dynamical systems and
  symmetry}}, Cambridge Monographs in Mechanics (\bibinfo{publisher}{Cambridge
  University Press}, \bibinfo{address}{Cambridge, England},
  \bibinfo{year}{2012}), \bibinfo{edition}{2nd} ed.

\bibitem[{\citenamefont{Manohar et~al.}(2017)\citenamefont{Manohar, Brunton,
  Kutz, and Brunton}}]{manohar2017data}
\bibinfo{author}{\bibfnamefont{K.}~\bibnamefont{Manohar}},
  \bibinfo{author}{\bibfnamefont{B.~W.} \bibnamefont{Brunton}},
  \bibinfo{author}{\bibfnamefont{J.~N.} \bibnamefont{Kutz}}, \bibnamefont{and}
  \bibinfo{author}{\bibfnamefont{S.~L.} \bibnamefont{Brunton}},
  \bibinfo{journal}{arXiv preprint arXiv:1701.07569}  (\bibinfo{year}{2017}).

\bibitem[{\citenamefont{Everson and Sirovich}(1995)}]{sirovich1995}
\bibinfo{author}{\bibfnamefont{R.}~\bibnamefont{Everson}} \bibnamefont{and}
  \bibinfo{author}{\bibfnamefont{L.}~\bibnamefont{Sirovich}},
  \bibinfo{journal}{J. Opt. Soc. Am. A} \textbf{\bibinfo{volume}{12}},
  \bibinfo{pages}{1657} (\bibinfo{year}{1995}).

\bibitem[{\citenamefont{Willcox}(2006)}]{willcox2005}
\bibinfo{author}{\bibfnamefont{K.}~\bibnamefont{Willcox}},
  \bibinfo{journal}{Computers \& Fluids} \textbf{\bibinfo{volume}{35}},
  \bibinfo{pages}{208} (\bibinfo{year}{2006}).

\bibitem[{\citenamefont{Yildirim et~al.}(2009)\citenamefont{Yildirim,
  Chryssostomidis, and Karniadakis}}]{Yildirim:2009}
\bibinfo{author}{\bibfnamefont{B.}~\bibnamefont{Yildirim}},
  \bibinfo{author}{\bibfnamefont{C.}~\bibnamefont{Chryssostomidis}},
  \bibnamefont{and} \bibinfo{author}{\bibfnamefont{G.~E.}
  \bibnamefont{Karniadakis}}, \bibinfo{journal}{Ocean Modelling}
  \textbf{\bibinfo{volume}{27}}, \bibinfo{pages}{160} (\bibinfo{year}{2009}).

\bibitem[{\citenamefont{Chaturantabut and Sorensen}(2010)}]{sorensen2010}
\bibinfo{author}{\bibfnamefont{S.}~\bibnamefont{Chaturantabut}}
  \bibnamefont{and} \bibinfo{author}{\bibfnamefont{D.~C.}
  \bibnamefont{Sorensen}}, \bibinfo{journal}{SIAM J. Sci. Comput.}
  \textbf{\bibinfo{volume}{32}}, \bibinfo{pages}{2737} (\bibinfo{year}{2010}).

\bibitem[{\citenamefont{Sargsyan et~al.}(2015)\citenamefont{Sargsyan, Brunton,
  and Kutz}}]{sargsyan2015}
\bibinfo{author}{\bibfnamefont{S.}~\bibnamefont{Sargsyan}},
  \bibinfo{author}{\bibfnamefont{S.~L.} \bibnamefont{Brunton}},
  \bibnamefont{and} \bibinfo{author}{\bibfnamefont{J.~N.} \bibnamefont{Kutz}},
  \bibinfo{journal}{Phys. Rev. E} \textbf{\bibinfo{volume}{92}},
  \bibinfo{pages}{033304} (\bibinfo{year}{2015}).

\bibitem[{\citenamefont{Schmid}(2010)}]{Schmid2010jfm}
\bibinfo{author}{\bibfnamefont{P.~J.} \bibnamefont{Schmid}},
  \bibinfo{journal}{Journal of Fluid Mechanics} \textbf{\bibinfo{volume}{656}},
  \bibinfo{pages}{5} (\bibinfo{year}{2010}), ISSN \bibinfo{issn}{0022-1120}.

\bibitem[{\citenamefont{Brunton et~al.}(2015)\citenamefont{Brunton, Proctor,
  Tu, and Kutz}}]{Brunton2015jcd}
\bibinfo{author}{\bibfnamefont{S.~L.} \bibnamefont{Brunton}},
  \bibinfo{author}{\bibfnamefont{J.~L.} \bibnamefont{Proctor}},
  \bibinfo{author}{\bibfnamefont{J.~H.} \bibnamefont{Tu}}, \bibnamefont{and}
  \bibinfo{author}{\bibfnamefont{J.~N.} \bibnamefont{Kutz}}
  (\bibinfo{year}{2015}), \bibinfo{note}{to appear in the Journal of
  Computational Dynamics. Available: arXiv:1312.5186}.

\bibitem[{\citenamefont{Tu et~al.}(2014{\natexlab{a}})\citenamefont{Tu, Rowley,
  Kutz, and Shang}}]{Tu:2014b}
\bibinfo{author}{\bibfnamefont{J.~H.} \bibnamefont{Tu}},
  \bibinfo{author}{\bibfnamefont{C.~W.} \bibnamefont{Rowley}},
  \bibinfo{author}{\bibfnamefont{J.~N.} \bibnamefont{Kutz}}, \bibnamefont{and}
  \bibinfo{author}{\bibfnamefont{J.~K.} \bibnamefont{Shang}},
  \bibinfo{journal}{Experiments in Fluids} \textbf{\bibinfo{volume}{55}},
  \bibinfo{pages}{1} (\bibinfo{year}{2014}{\natexlab{a}}).

\bibitem[{\citenamefont{Askham and Kutz}(2018)}]{askham2018variable}
\bibinfo{author}{\bibfnamefont{T.}~\bibnamefont{Askham}} \bibnamefont{and}
  \bibinfo{author}{\bibfnamefont{J.~N.} \bibnamefont{Kutz}},
  \bibinfo{journal}{SIAM Journal on Applied Dynamical Systems}
  \textbf{\bibinfo{volume}{17}}, \bibinfo{pages}{380} (\bibinfo{year}{2018}).

\bibitem[{\citenamefont{Tu et~al.}(2014{\natexlab{b}})\citenamefont{Tu, Rowley,
  Luchtenburg, Brunton, and Kutz}}]{Tu2014jcd}
\bibinfo{author}{\bibfnamefont{J.~H.} \bibnamefont{Tu}},
  \bibinfo{author}{\bibfnamefont{C.~W.} \bibnamefont{Rowley}},
  \bibinfo{author}{\bibfnamefont{D.~M.} \bibnamefont{Luchtenburg}},
  \bibinfo{author}{\bibfnamefont{S.~L.} \bibnamefont{Brunton}},
  \bibnamefont{and} \bibinfo{author}{\bibfnamefont{J.~N.} \bibnamefont{Kutz}},
  \bibinfo{journal}{Journal of Computational Dynamics}
  \textbf{\bibinfo{volume}{1}}, \bibinfo{pages}{391}
  (\bibinfo{year}{2014}{\natexlab{b}}).

\bibitem[{\citenamefont{Mezic}(2013)}]{mezic2013analysis}
\bibinfo{author}{\bibfnamefont{I.}~\bibnamefont{Mezic}},
  \bibinfo{journal}{Annual Review of Fluid Mechanics}
  \textbf{\bibinfo{volume}{45}}, \bibinfo{pages}{357} (\bibinfo{year}{2013}).

\bibitem[{\citenamefont{Kutz et~al.}(2016)\citenamefont{Kutz, Brunton, Brunton,
  and Proctor}}]{DMDbook}
\bibinfo{author}{\bibfnamefont{J.~N.} \bibnamefont{Kutz}},
  \bibinfo{author}{\bibfnamefont{S.~L.} \bibnamefont{Brunton}},
  \bibinfo{author}{\bibfnamefont{B.~W.} \bibnamefont{Brunton}},
  \bibnamefont{and} \bibinfo{author}{\bibfnamefont{J.~L.}
  \bibnamefont{Proctor}}, \emph{\bibinfo{title}{Dynamic mode decomposition:
  Data-driven modeling of complex systems}} (\bibinfo{year}{2016}).

\bibitem[{\citenamefont{Bright et~al.}(2013)\citenamefont{Bright, Lin, and
  Kutz}}]{bright2013}
\bibinfo{author}{\bibfnamefont{I.}~\bibnamefont{Bright}},
  \bibinfo{author}{\bibfnamefont{G.}~\bibnamefont{Lin}}, \bibnamefont{and}
  \bibinfo{author}{\bibfnamefont{J.~N.} \bibnamefont{Kutz}},
  \bibinfo{journal}{Phys. Fluids} \textbf{\bibinfo{volume}{25}},
  \bibinfo{pages}{127102} (\bibinfo{year}{2013}).

\bibitem[{\citenamefont{Brunton et~al.}(2014)\citenamefont{Brunton, Tu, Bright,
  and Kutz}}]{Brunton2014siads}
\bibinfo{author}{\bibfnamefont{S.~L.} \bibnamefont{Brunton}},
  \bibinfo{author}{\bibfnamefont{J.~H.} \bibnamefont{Tu}},
  \bibinfo{author}{\bibfnamefont{I.}~\bibnamefont{Bright}}, \bibnamefont{and}
  \bibinfo{author}{\bibfnamefont{J.~N.} \bibnamefont{Kutz}},
  \bibinfo{journal}{SIAM Journal on Applied Dynamical Systems}
  \textbf{\bibinfo{volume}{13}}, \bibinfo{pages}{1716} (\bibinfo{year}{2014}).

\bibitem[{\citenamefont{Proctor et~al.}(2014)\citenamefont{Proctor, Brunton,
  Brunton, and Kutz}}]{Proctor2014epj}
\bibinfo{author}{\bibfnamefont{J.~L.} \bibnamefont{Proctor}},
  \bibinfo{author}{\bibfnamefont{S.~L.} \bibnamefont{Brunton}},
  \bibinfo{author}{\bibfnamefont{B.~W.} \bibnamefont{Brunton}},
  \bibnamefont{and} \bibinfo{author}{\bibfnamefont{J.~N.} \bibnamefont{Kutz}},
  \bibinfo{journal}{The European Physical Journal Special Topics}
  \textbf{\bibinfo{volume}{223}}, \bibinfo{pages}{2665} (\bibinfo{year}{2014}).

\bibitem[{\citenamefont{Kramer et~al.}(2015)\citenamefont{Kramer, Grover,
  Boufounos, Benosman, and Nabi}}]{kramer}
\bibinfo{author}{\bibfnamefont{B.}~\bibnamefont{Kramer}},
  \bibinfo{author}{\bibfnamefont{P.}~\bibnamefont{Grover}},
  \bibinfo{author}{\bibfnamefont{P.}~\bibnamefont{Boufounos}},
  \bibinfo{author}{\bibfnamefont{M.}~\bibnamefont{Benosman}}, \bibnamefont{and}
  \bibinfo{author}{\bibfnamefont{S.}~\bibnamefont{Nabi}},
  \bibinfo{journal}{arXiv preprint arXiv:1510.02831}  (\bibinfo{year}{2015}).

\bibitem[{\citenamefont{Cand\`es et~al.}(59)\citenamefont{Cand\`es, Romberg,
  and Tao}}]{Candes:2006c}
\bibinfo{author}{\bibfnamefont{E.~J.} \bibnamefont{Cand\`es}},
  \bibinfo{author}{\bibfnamefont{J.}~\bibnamefont{Romberg}}, \bibnamefont{and}
  \bibinfo{author}{\bibfnamefont{T.}~\bibnamefont{Tao}},
  \bibinfo{journal}{Communications in Pure and Applied Mathematics}
  \textbf{\bibinfo{volume}{8}} (\bibinfo{year}{59}).

\bibitem[{\citenamefont{Donoho}(2006)}]{Donoho:2006}
\bibinfo{author}{\bibfnamefont{D.~L.} \bibnamefont{Donoho}},
  \bibinfo{journal}{IEEE Transactions on Information Theory}
  \textbf{\bibinfo{volume}{52}}, \bibinfo{pages}{1289} (\bibinfo{year}{2006}).

\bibitem[{\citenamefont{Tropp and Gilbert}(2007)}]{tropp2007signal}
\bibinfo{author}{\bibfnamefont{J.~A.} \bibnamefont{Tropp}} \bibnamefont{and}
  \bibinfo{author}{\bibfnamefont{A.~C.} \bibnamefont{Gilbert}},
  \bibinfo{journal}{IEEE Transactions on information theory}
  \textbf{\bibinfo{volume}{53}}, \bibinfo{pages}{4655} (\bibinfo{year}{2007}).

\bibitem[{\citenamefont{Cand\`es}(2006)}]{Candes:2006}
\bibinfo{author}{\bibfnamefont{E.~J.} \bibnamefont{Cand\`es}},
  \bibinfo{journal}{Proceedings of the International Congress of Mathematics}
  (\bibinfo{year}{2006}).

\bibitem[{\citenamefont{Cand\`es et~al.}(2006)\citenamefont{Cand\`es, Romberg,
  and Tao}}]{Candes:2006a}
\bibinfo{author}{\bibfnamefont{E.~J.} \bibnamefont{Cand\`es}},
  \bibinfo{author}{\bibfnamefont{J.}~\bibnamefont{Romberg}}, \bibnamefont{and}
  \bibinfo{author}{\bibfnamefont{T.}~\bibnamefont{Tao}}, \bibinfo{journal}{IEEE
  Transactions on Information Theory} \textbf{\bibinfo{volume}{52}},
  \bibinfo{pages}{489} (\bibinfo{year}{2006}).

\bibitem[{\citenamefont{Baraniuk}(2007)}]{Baraniuk:2007}
\bibinfo{author}{\bibfnamefont{R.~G.} \bibnamefont{Baraniuk}},
  \bibinfo{journal}{IEEE Signal Processing Magazine}
  \textbf{\bibinfo{volume}{24}}, \bibinfo{pages}{118} (\bibinfo{year}{2007}).

\bibitem[{\citenamefont{Cand\`es and Tao}(2006)}]{Candes:2006b}
\bibinfo{author}{\bibfnamefont{E.~J.} \bibnamefont{Cand\`es}} \bibnamefont{and}
  \bibinfo{author}{\bibfnamefont{T.}~\bibnamefont{Tao}}, \bibinfo{journal}{IEEE
  Transactions on Information Theory} \textbf{\bibinfo{volume}{52}},
  \bibinfo{pages}{5406} (\bibinfo{year}{2006}).

\bibitem[{\citenamefont{Rogers}(2012)}]{rogers2012power}
\bibinfo{author}{\bibfnamefont{G.}~\bibnamefont{Rogers}},
  \emph{\bibinfo{title}{Power system oscillations}}
  (\bibinfo{publisher}{Springer Science \& Business Media},
  \bibinfo{year}{2012}).

\end{thebibliography}

\end{document}